\documentstyle[aps,prl,twocolumn,epsfig,floats,amssymb,amsfonts]{revtex}

\newcommand{\mb}[1]{\mbox{\boldmath $#1$}}

\begin{document}
 
\draft
\title{Enhanced mesoscopic fluctuations in the crossover between random matrix
ensembles}

\author{Shaffique Adam, Piet W.\ Brouwer, James P.\ Sethna, Xavier Waintal\cite{present}}

\address{
{}Laboratory of Atomic and Solid State Physics,
Cornell University, Ithaca, NY 14853-2501}

\twocolumn[
\widetext
\begin{@twocolumnfalse}

\maketitle

\begin{abstract}
In random-matrix ensembles that interpolate between the three basic 
ensembles (orthogonal, unitary, and symplectic), there exist
correlations between elements of the same eigenvector and between
different eigenvectors. We study such correlations, using a remarkable 
correspondence between
the interpolating ensembles late in the crossover and a basic
ensemble of finite size. In small metal grains or
semiconductor quantum dots, the correlations between different eigenvectors
lead to enhanced fluctuations of the electron-electron interaction matrix 
elements which become parametrically larger than the non-universal 
fluctuations.
\end{abstract}


\end{@twocolumnfalse}
]

\narrowtext
Random matrix theory has focused on the study of three ensembles 
of Hamiltonians: the Gaussian Unitary Ensemble (GUE), the Gaussian Orthogonal 
Ensemble (GOE), and the Gaussian Symplectic Ensemble (GSE). These describe
the statistics of single-particle energy levels and wavefunctions of 
disordered metal grains or chaotic quantum dots with the corresponding
symmetries; GUE if time-reversal symmetry is broken, and GOE or GSE if
time-reversal symmetry is present and spin-rotation symmetry 
is present or absent, respectively. In these three basic ensembles,
eigenvector elements are Gaussian complex/real/quaternion random numbers;
elements of the same eigenvector and of different eigenvectors are
all statistically independent \cite{Mehta}.

Disordered or chaotic systems with partially broken symmetries 
show a variety of phenomena
that go beyond a mere ``interpolation'' of descriptions
based on the GOE, GUE, and GSE alone. 
For example, in a quantum dot, a weak magnetic field causes 
long-range wavefunction
correlations \cite{French,Sommers,Falko} and a
non-Gaussian distribution of ``level velocities'', derivatives
of energy levels with respect to, e.g., a shape change of the dot
\cite{VanLangen}. Both effects are absent without
a magnetic field (in the GOE), or when the magnetic field is strong 
enough to fully break time-reversal symmetry (in the GUE). 
In a metal grain, weak spin-orbit interaction induces mesoscopic
fluctuations of the $g$-tensor \cite{BWH,Matveev}, which does not
fluctuate in either the GOE or the GSE. 
Further, as we'll show below, in a 
weak magnetic field or for weak spin-orbit scattering,
matrix elements of the electron-electron
interaction exhibit fluctuations that are parametrically larger
than in each of the three basic ensembles.

The underlying reason for these phenomena is that eigenvector
elements are not independent in (random-matrix) ensembles that
interpolate between the three basic symmetry classes: There exist both
correlations within the same eigenvector 
\cite{French,Sommers,Falko,VanLangen,BWH}
and, as we show in this letter, between different eigenvectors. 
To study the eigenvector correlations in such crossover ensembles, we 
will make use of a surprising relation
between the eigenvector statistics late in the crossover from class 
A to class B and that of finite-sized matrices in class B (where B is 
the 
class of lower symmetry). Examples of such a relation were known for
the statistics of a single eigenvector. For example, in the GOE-GUE
crossover, which is described by the $N \times N$
random hermitian matrix (with $N$ taken to $\infty$ at the end of
the calculation)
\cite{Pandey}
\begin{equation}
  H_{\rm OU}(N,\alpha) = H_{\rm O}(N) + {\alpha \over
\sqrt{N}} H_{\rm U}(N),
  \label{eq:HGOEGUE}
\end{equation}
the distribution of the ``phase rigidity'' $|{\mb{v}}^{\rm T} {\mb{v}}|^2$ 
\cite{VanLangen} of a single eigenvector ${\mb{v}}$ is the same as in
the {\em finite-sized} $M \times M$ GUE ensemble with $M=2\alpha^2$
if $\alpha$ is large. In Eq.\ (\ref{eq:HGOEGUE}), $H_{\rm O}(N)$ and
$H_{\rm U}(N)$ are $N \times N$ matrices
taken from the GOE and GUE, respectively, with equal variances for
the 
%individual 
matrix elements. 
A similar correspondence occurs for the
$g$-tensor of a Kramers doublet in the GOE-GSE crossover 
\cite{BWH,Matveev}. 
Our main finding is
that such a correspondence extends to the
correlations between different eigenvectors.

In this paper we will accomplish four tasks.
(i)~We show numerically that the relation
\begin{equation}
  \label{mainres}
  H_{\rm OU}(N,\alpha) \leftrightarrow
  H_{\rm U}(M),\ \ M = 2 \alpha^2
\end{equation}
between the GOE-GUE crossover
Hamiltonian $H_{\rm OU}(\alpha)$ for large
$\alpha$ and $N$
and a finite-sized
$M \times M$ GUE Hamiltonian extends to correlations between 
eigenvectors. 
Just as in critical phenomena, where simple power laws unfold into
universal scaling functions as you flow away from the critical point,
here a rich theory of correlations unfolds in the crossover region.
We wish to point out that this principle applies not only to the 
GOE-GUE crossover, but also, e.g., to the GOE-GSE crossover,
% which describes metal grains with spin-orbit scattering, 
or to wavefunctions in two coupled quantum dots, which are
described by a random Hamiltonian interpolating between two
independent GUE's and one GUE of double size \cite{Tschersich}.
(ii)~We show that, for large $\alpha$,
the universality classes are actually curves in the $(1/\alpha,1/N)$
plane, reminiscent of renormalization-group flow trajectories
\cite{renormalization}. 
(iii)~We calculate correlations between eigenvectors, based on the 
surmise (\ref{mainres})
and diagrammatic perturbation theory.
(iv)~We calculate how the inter-eigenvector 
correlations in the crossover region affect matrix elements of the 
electron-electron interaction in a quantum dot or metal grain in a
weak magnetic field, and predict a significant enhancement of
fluctuations compared to the basic ensembles.

Let us now consider the joint distribution $P(\{\mb{v}_{\mu}\})$
of $n$ eigenvectors 
${\mb{v}}_{\mu}$, $\mu=1,\ldots,n$, for the example of the GOE-GUE crossover 
Hamiltonian (\ref{eq:HGOEGUE}). 
Throughout the entire GOE-GUE crossover, the distribution of the
eigenvectors is invariant under orthogonal transformations.
As a consequence, the joint distribution $P(\{{\mb{v}}_{\mu}\})$
is completely determined by the distribution of 
the orthogonal invariants \cite{French,Sommers}
\begin{equation}
  \rho_{\mu \nu} = \rho_{\nu \mu} =
  {\mb{v}}^{\rm T}_{\mu} {\mb{v}}^{\vphantom{T}}_{\nu},\ \
  \mu, \nu = 1,\ldots,n,
\end{equation}
where the superscript ${\rm T}$ denotes transposition. Hence
\begin{eqnarray}
  P(\{ {\mb{v}}_{\mu} \}) &=&
  \int \prod_{\mu \le \nu}^n d\rho_{\mu\nu}\, P(\{\rho_{\mu\nu}\})
  \nonumber \\ && \mbox{} \times
  \prod_{\mu \le \nu}^n \delta({\mb{v}}_{\mu}^{\dagger} {\mb{v}}_{\nu} -
  \delta_{\mu\nu}) \delta({\mb{v}}_{\mu}^{\rm T} {\mb{v}}_{\nu} -
  \rho_{\mu\nu}).
  \label{eq:Pv} 
\end{eqnarray}
For the physically relevant case of large $N$, Eq.\
(\ref{eq:Pv}) implies that the eigenvector
elements ${\mb{v}}_{\mu m}$, $m=1,\ldots,N$,
have a Gaussian distribution with zero mean and
\begin{equation}
  \langle v_{\mu m}^* v_{\nu n}
  \rangle_{\rho}
  = N^{-1} \delta_{\mu \nu} \delta_{mn}, \ \
  \langle v_{\mu m} v_{\nu n} \rangle_{\rho} = 
  \rho_{\mu \nu} \delta_{mn}.
  \label{eq:vvavg}
\end{equation}
The subscript $\langle \ldots \rangle_{\rho}$ indicates
that the average is taken at fixed $\rho_{\mu \nu}$.
For the full ensemble
average one has to perform a subsequent average over the
$\rho_{\mu\nu}$ with the distribution $P(\{\rho_{\mu\nu}\})$.
We can find $P(\{\rho_{\mu\nu}\})$ from the surmise that, for 
$\alpha \gg 1$ and for eigenvectors ${\mb{v}}_{\mu}$ whose energies are all
inside a window of size $\lesssim \alpha^2 \Delta$, $\Delta$
being the level spacing of the Hamiltonian $H(\alpha)$, the
joint distribution of the $\rho_{\mu \nu}$ {\em is the same
as for a GUE Hamiltonian of finite size} $M = 2 \alpha^2$.
Thus the $\rho_{\mu \nu}$ are independently and Gaussian
distributed with zero mean and with variance
\begin{equation}
  \langle |\rho_{\mu \nu}|^2 \rangle
  = (1 + \delta_{\mu \nu})/M,\ \ M = 2\alpha^2.
  \label{eq:Pa}
\end{equation}
Together, Eqs.\ (\ref{eq:Pv})--(\ref{eq:Pa}) fix the joint
distribution of eigenvectors in the crossover ensemble close to
the GUE. For the single-eigenvector distribution, 
Eqs.\ (\ref{eq:Pv})--(\ref{eq:Pa}) reproduce 
the $\alpha \gg 1$ limit of the
exact solution of Sommers and Iida \cite{Sommers}.
The fact that the phase rigidity $|\rho_{\mu \mu}|^2$ of a
single eigenvector is a fluctuating quantity is the prime
cause of the correlations between elements of one eigenvector
\cite{Falko,VanLangen};
It is the existence of nonzero and fluctuating $\rho_{\mu \nu}$ 
for $\mu \neq \nu$ that causes the correlations between different 
eigenvectors.

We now proceed to present arguments in support of our surmise.
We consider the eigenvectors ${\mb{v}}_{\mu}$ ($\mu=1,\ldots,n$)
with energies within a distance $\ll \alpha^2 \Delta$ 
from a reference energy $\varepsilon_{\rm ref}$, sorting them by 
increasing energy. We then
consider how each of these eigenvectors is
built up from the eigenvectors ${\mb{o}}_{\nu}$ of the unperturbed 
Hamiltonian $H_{\rm O}$. Contributions from eigenvectors ${\mb{o}}_{\nu}$ 
with energy $\varepsilon_{\nu}$
far away from $\varepsilon_{\rm ref}$ can be described using perturbation
theory in the crossover parameter $\alpha$.
If the energy difference $\varepsilon_{\rm ref} - \varepsilon_{\nu}$
is sufficiently large, the admixture of ${\mb{o}}_{\nu}$ to 
any of the vectors ${\mb{v}}_{\mu}(\alpha)$ of interest
is small and can be neglected.
On the other hand, eigenvectors ${\mb{o}}_{\nu}$
with energy $\varepsilon_{\nu}$
close to $\varepsilon_{\rm ref}$ contribute non-perturbatively for
large $\alpha$. Upon increasing $\alpha$ from zero, the eigenvectors
${\mb{v}}_{\mu}(\alpha)$ in this energy range
have undergone several avoided crossings, and the unperturbed
eigenvectors ${\mb{o}}_{\mu}$ have roughly equal weights
in each of the vectors ${\mb{v}}_{\mu}(\alpha)$ in our set.

\begin{figure}
\bigskip
\epsfxsize=0.8\hsize
\hspace{0.05\hsize}
\epsffile{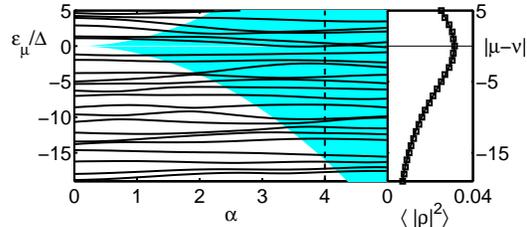}
\caption{\label{fig:2} Left panel: Eigenvalues for one realization of
$H_{\rm OU}(\alpha)$. The shaded region marks the energy window
of size $\sim M(\alpha) \Delta = 2 \alpha^2 \Delta$ for which the 
eigenvalues are kept in the effective $M \times M$ GUE Hamiltonian.
Right panel: $\langle |\rho_{\mu \nu}|^2
\rangle$ as a function of the distance $\mu - \nu \approx
(\varepsilon_{\mu} - \varepsilon_{\nu})/\Delta$ between eigenvalues,
for $\alpha = 4.0$ (dashed line, left panel).
Solid curve: Eq.\ (\ref{eq:rhoMuNu}). Data points: numerical
calculation for $N=400$.\vspace{-0.3cm}}
\end{figure}

It is on this heuristic picture that our surmise for 
an effective description of the eigenvector statistics for large 
$\alpha$ is based: 
We only retain those eigenvectors of the unperturbed Hamiltonian 
$H_{\rm O}$ that are relatively close in energy and hence all
contribute roughly equally, see Fig.\ \ref{fig:2} for a cartoon. 
Since the time-reversal symmetry
breaking perturbation in Eq.\ (\ref{eq:HGOEGUE})
is strong for these eigenvectors, the 
matrix elements between them form a random hermitian
matrix of the GUE. Denoting the 
effective number of contributing unperturbed eigenvectors as 
$M(\alpha)$, we thus reduce the problem of finding the
distribution of the orthogonal invariants $\rho_{\mu\nu}$ for
the $N \times N$ crossover Hamiltonian (\ref{eq:HGOEGUE}) to that 
of finding the distribution of the $\rho_{\mu \nu}$
for the much smaller GUE Hamiltonian of size $M(\alpha)$. 
To calculate
$M(\alpha)$ in terms of $N$ and $\alpha$, we turn to the
exact solution for the single-eigenvector
distribution obtained in Refs.\ \onlinecite{Sommers,Falko,VanLangen},
and find \cite{foot}
\begin{equation}
  \label{eq:MN}
  M(\alpha) = \alpha^2 N (\alpha^2 + 2 N)/(\alpha^2+N)^2.
\end{equation}
For large $N$ this simplifies to $M(\alpha) = 2 \alpha^2$, 
in agreement with Eq.\ (\ref{eq:Pa}).

\begin{figure}
\bigskip
\epsfxsize=0.74\hsize
\hspace{0.08\hsize}
\epsffile{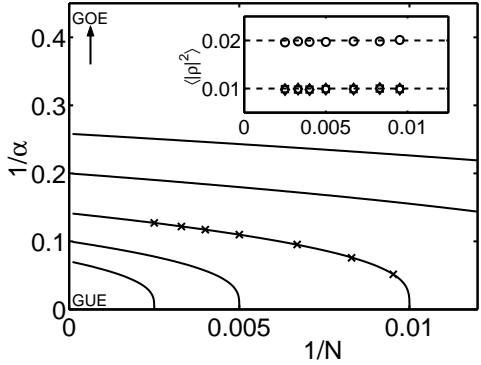}
\caption{\label{fig:1} Curves of constant effective GUE size
$M(\alpha)$, Eq.\ (\protect\ref{eq:MN}),
in the $(1/\alpha,1/N)$ plane for the $N \times N$
crossover Hamiltonian (\protect\ref{eq:HGOEGUE}). Top to
bottom: $M=20$, $M=30$, $M=50$, $M=100$, and $M=200$. 
The horizontal and vertical axes correspond to the pure GUE
and to the $N\to\infty$ crossover Hamiltonian, respectively.
Inset: 
$\langle |\rho_{\mu \nu}|^2 \rangle$ for the points indicated
at the $M=50$ curve in the main panel. 
Circles: $\mu = \nu$; Squares: $\mu = \nu + 1$ 
(eigenvectors with neighboring energy levels); diamonds:
$\mu = \nu + 2$ (next-nearest neighbors). The dashed lines 
indicate the surmise of Eq.\ (\ref{eq:Pa}).\vspace{-0.3cm}}
\end{figure}

By our surmise, the distribution of the orthogonal invariants should
depend on the effective matrix size
$M(\alpha)$ only, not on $\alpha$ and $N$ individually,
as long as $N$ and $\alpha$ are large. We have verified this 
by numerical calculation of the averages
$\langle |\rho_{\mu \nu}|^2 \rangle$ for different points along a
curve of constant $M(\alpha)$ in the $(1/N,1/\alpha)$ plane. 
The results of such a calculation are shown in Fig.\ \ref{fig:1}
for $\mu = \nu$, for neighboring eigenvectors ($\mu = \nu+1$) and 
for next-nearest neighbors ($\mu = \nu + 2$).
We have also verified that the
distribution of the $\rho_{\mu \nu}$ is indeed Gaussian (not
shown).

The surmise (\ref{mainres}) is expected to be valid as long 
as only eigenvectors taken from an energy window of width 
$\ll M(\alpha) \Delta = 2 \alpha^2 \Delta$ are involved. If the
energy differences between eigenvectors become of order $\alpha^2
\Delta$ or larger, the 
eigenvectors ${\mb{v}}_{\mu}(\alpha)$ do not share 
the same unperturbed eigenvectors ${\mb{o}}_{\nu}$, and we thus
expect that they become uncorrelated. A quantitative description 
of eigenvector correlations at energy separations 
$\gg \Delta$ can be 
obtained using diagrammatic perturbation theory. 
The only nonzero second moment is $\langle
|\rho_{\mu \nu}|^2 \rangle$, which can be computed from
\begin{eqnarray}
  \langle |\rho_{\mu\nu}|^2 \rangle &=&
  -{\Delta^2 \over 4 \pi^2}
  \sum_{s_1, s_2 = \pm}
  s_1 s_2 
  \nonumber \\ && \mbox{} \times 
  \left\langle \mbox{tr}\,  G^{\rm T}(\varepsilon_{\mu} + i s_1 \delta)
  G(\varepsilon_{\nu} + is_2 \delta) \right\rangle.
\end{eqnarray}
where $G(z) = 1/(z - H_{\rm OU})$, $\delta$ is a positive
 infinitesimal, and
the eigenvectors ${\mb{v}}_{\mu}$ and ${\mb{v}}_{\nu}$ have energies 
$\varepsilon_{\mu}$ and $\varepsilon_{\nu}$, respectively. 
Calculating the averages using the technique of Ref.\ \onlinecite{BrezinZee}, 
we find, if $\mu \neq \nu$,
\begin{equation}
  \langle |\rho_{\mu \nu}|^2 \rangle =
  {2 \alpha^2 \over
   4 \alpha^4 + \pi^2 (\varepsilon_{\mu} -
  \varepsilon_{\nu})^2/\Delta^2}.
  \label{eq:rhoMuNu}
\end{equation} 
A similar result for parametric correlations inside
a basic random-matrix ensemble was derived in Ref.\
\onlinecite{Wilkinson}.
%Upon increasing $\alpha$ the number of correlated
%eigenvectors increases proportional to $\alpha^2$, while the 
%degree of correlation decreases with the same rate.
The right panel of Fig.\ \ref{fig:2} shows  $\langle |\rho_{\mu \nu}|^2
\rangle$ as a function of $\varepsilon_{\mu} - \varepsilon_{\nu}$
and a numerical calculation of the same quantity.

The GOE-GUE crossover that we considered here describes the
wavefunction statistics in, e.g., a chaotic quantum dot or a
disordered metal grain in a weak magnetic field.
Wavefunction distributions have immediate experimental relevance
for the spacings, widths, and heights of Coulomb blockade peaks
in the conductance of metal grains or quantum dots \cite{ABG}.
Correlations between wavefunctions of neighboring energy levels
cause correlations between the heights and widths of neighboring 
conductance peaks. Wavefunction distributions also
influence the positions of Coulomb blockade peaks through their role
in the distribution of electron-electron interaction matrix elements
\cite{vonDelftRalph}, which we now discuss in detail.
The interaction matrix element $U_{\mu\nu\rho\sigma}$ is defined as
\begin{eqnarray}
  U_{\mu\nu\rho\sigma} &=&
  \int d\vec{r}_1 d\vec{r}_2 U(\vec{r}_1 - \vec{r}_2) \nonumber \\
  && \mbox{} \times
  \phi_{\mu}(\vec{r}_1) \phi_{\nu}(\vec{r}_2)
  \phi_{\rho}(\vec{r}_2)^* \phi_{\sigma}(\vec{r}_1)^*,
\end{eqnarray}
where $U(\vec{r})$ is the electron-electron interaction potential and
$\phi_{\mu}(\vec{r})$ the wavefunction for an electron in
level $\varepsilon_{\mu}$. For example, the difference of
interaction matrix elements $U_{\mu \mu \nu \nu} - U_{\mu \mu
oo}$ gives the spacing between peak positions corresponding to
different nonequilibrium configurations (levels $\nu$ and $o$
unoccupied, respectively) in tunneling spectroscopy
of small metal grains \cite{Agam}.

In a metal grain or quantum dot, the
interaction can be approximated by an $\vec{r}$-independent
part and a local interaction
$U^{\rm loc}(\vec{r}) = \lambda \Delta V \delta(\vec r)$, where
$\Delta$ is the mean level spacing, $V$ is the sample volume, and
$\lambda$ is a parameter of order unity
governing the strength of the local
interaction. The spatially constant interaction leads to a
charging energy and does not show mesoscopic fluctuations. For
the GOE (no magnetic field), the
matrix elements of $U^{\rm loc}$ have an average given 
by \cite{ABG}
\begin{eqnarray}
  \langle U^{\rm loc}_{\mu\nu\rho\sigma}\rangle =
  \lambda \Delta (\delta_{\mu \sigma} \delta_{\nu\rho}
  + \delta_{\mu \rho} \delta_{\nu \sigma} + \delta_{\mu\nu}
  \delta_{\rho\sigma}).
  \label{eq:Uavg}
\end{eqnarray}
If time-reversal symmetry is broken by a magnetic field (i.e., in
the GUE), the last term in Eq.\ (\ref{eq:Uavg}) is left out 
\cite{renorm}.
In both the GOE and GUE, 
fluctuations of the interaction matrix elements 
$U^{\rm loc}_{\mu\nu\rho\sigma}$
and corrections to Eq.\ (\ref{eq:Uavg}) are nonuniversal
and small as (at most) $g^{-1/2}$, where
$g$ is the dimensionless conductance of the sample. 
Equation (\ref{eq:Uavg}) can be reproduced from
random-matrix theory if the wavefunctions $\phi_{\mu}(\vec{r})$ are 
replaced by
eigenvectors ${\mb{v}}_{\mu}$ and the integration over space is replaced by
a summation over the vector indices.

\begin{figure}
\epsfxsize=0.8\hsize
\hspace{0.1\hsize}
\epsffile{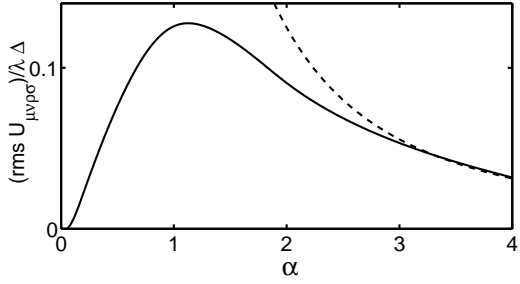}

\caption{\label{fig:3} Root-mean-square fluctuations of the
interaction matrix element $U_{\mu\nu\rho\sigma}$ for four
consecutive levels $\mu=\nu-1$ $=\rho-2=\sigma-3$.
The dashed line shows the large-$\alpha$ asymptote of Eq.\ 
(\protect\ref{eq:Uvar}). The solid line is obtained from
numerical generation of $400 \times 400$ GOE-GUE crossover 
matrices, using
$\langle |U_{\rm \mu\nu\rho\sigma} |^2 \rangle =
  (\lambda \Delta)^2 \langle |\rho_{\mu \nu}|^2 |\rho_{\rho
\sigma}|^2 \rangle$. (Direct numerical calculation of 
$U_{\rm \mu\nu\rho\sigma}$ suffers from large finite-$N$
corrections.)\vspace{-0.3cm}
}
\end{figure}

How are the interaction matrix elements distributed in the presence
of a weak magnetic field? 
If we are not interested in the non-universal
($1/g$) corrections, that question can be answered using the
eigenvector distributions for the GOE-GUE crossover
that we derived above. 
First, upon increasing the magnetic field,
there is a gradual suppression of the last term 
in Eq.\ (\ref{eq:Uavg}). However, as a result of gauge invariance
(wave-functions can be multiplied with an arbitrary phase factor),
the {\em average} of the last term in Eq.\ (\ref{eq:Uavg}) is 
formally zero throughout the crossover if $\mu \neq \rho$, 
and this effect only shows
up in the fluctuations of $U_{\mu\nu\rho\sigma}$. Second, the
appearance of inter-eigenvector correlations enhances the
average of all interaction matrix elements. 
Using Eq.\ (\ref{eq:vvavg}), we find
%Averaging
%$U_{\mu\nu\rho\sigma}$ with the help of Eq.\ (\ref{eq:vvavg}),
%we find
\begin{eqnarray}
  \langle U_{\mu\nu\rho\sigma}^{\rm loc}\rangle &=&
  \lambda \Delta (\delta_{\mu \rho} \delta_{\nu\sigma} +
  \delta_{\mu \sigma} \delta_{\nu \rho} +
  \langle \rho_{\mu \nu} \rho_{\rho\sigma}\rangle).
\end{eqnarray}
For large $\alpha$, the average $\langle \rho_{\mu \nu}
\rho_{\rho\sigma} \rangle$ can be done with the help of
Eqs.\ (\ref{eq:Pa}) and (\ref{eq:rhoMuNu}), with the result
\begin{eqnarray}
  \langle U_{\mu\nu\rho\sigma}^{\rm loc}\rangle &=&
  \lambda \Delta  (\delta_{\mu \rho} \delta_{\nu\sigma} +
  \delta_{\mu \sigma} \delta_{\nu \rho}) \nonumber \\
  && \mbox{} \times
  \left( 1 + {2 \alpha^2  \over
   4 \alpha^4 + \pi^2 (\varepsilon_{\mu} -
  \varepsilon_{\nu})^2/\Delta^2} \right).
  \label{eq:Uavg2}
\end{eqnarray}
Third, the inter-eigenvector correlations enhance the fluctuations
of the interaction matrix elements. This is best illustrated by the
expectation value $\langle |U_{\mu\nu\rho\sigma}|^2\rangle$ with
all four indices $\mu$, $\nu$, $\rho$, and $\sigma$ different,
\begin{eqnarray}
  \langle |U_{\mu\nu\rho\sigma}^{\rm loc}|^2\rangle &=&
  (\lambda \Delta)^2
  \langle |\rho_{\mu \nu}|^2 |\rho_{\rho\sigma}|^2 \rangle
  = (\lambda \Delta)^2/(2 \alpha^2)^2.
  \label{eq:Uvar}
\end{eqnarray}
The first equality in Eq.\ (\ref{eq:Uvar}) holds for all
$\alpha$, the second one only if $\alpha \gg 1$ and the four
eigenvalues $\varepsilon_{\mu}$, $\varepsilon_{\nu}$, 
$\varepsilon_{\rho}$, $\varepsilon_{\sigma}$ are within a
distance $\ll \alpha^2 \Delta$ of each other. We have
numerically calculated $\langle |U_{\mu\nu\rho\sigma}^{\rm
loc}|^2\rangle$ for four neighboring wavefunctions
as a function of $\alpha$ throughout the entire GOE-GUE crossover,
see Fig.\ \ref{fig:3}.

Note that the enhanced fluctuations appear in  matrix elements
that are
{\em not} commonly associated with the presence of time-reversal
symmetry.
% (Cooperon terms).
Although the fluctuations are small if $\alpha \gg 1$,
%in the regime of large $\alpha$
%(i.e., far in the GOE-GUE crossover) for which they were derived, 
they
can be significantly larger than the non-universal fluctuations that
vanish as 
%$g^{-1}$ [for Eq.\ (\ref{eq:Uavg2})] or 
$g^{-2}$ [for Eq.\ (\ref{eq:Uvar})]. 
These extra large fluctuations can be an
additional source of fluctuations of Coulomb-blockade peak positions, 
and may dominate over the non-universal sources of fluctuations.

The origin of the eigenvector correlations and the
enhanced fluctuations of interaction matrix
elements can be sought in the existence of the large
parameter $\alpha^2$ that plays a role similar to the 
dimensionless conductance $g$ in the pure ensembles. 
The parameter $\alpha^2$ can be identified
as the ratio of the Heisenberg time $\tau_{\rm H} = 2 \pi
\hbar/\Delta$ and the time $\tau_{\rm OU}$ needed to acquire
a flux quantum \cite{ABG}.
Late in the crossover, GUE physics ranges from
the mean level spacing $\Delta$ up to the scale 
$\hbar/\tau_{\rm OU}$. 
In the pure GUE, however, validity of 
random-matrix theory ceases only at the higher energy scale 
$\hbar/\tau_{\rm erg}$, where $\tau_{\rm erg}$ is the ergodic
time. The role of the large parameter $g=\tau_{\rm H}/\tau_{\rm erg}$, 
which governs wavefunction
correlations and interaction matrix element fluctuations in the
``pure'' GUE and GOE is thus played by $\alpha^2 \sim \tau_{\rm
H}/\tau_{\rm OH}$ in the GOE-GUE crossover.

We thank Yuval Oreg and Dan Ralph for discussions.
This work was supported in part by the NSF under grants no.\ DMR
0086509 and KDI 9873214 and by the Sloan and Packard foundations.


\begin{references}
\bibitem{foot}
A rough estimate of $M$ can be obtained by comparing the
contributions to ${\mb{v}}_{\mu}(\alpha)$ from unperturbed
eigenvectors ${\mb{o}}_{\nu}$ with energy $\varepsilon_{\nu}$
close to (far away) from $\varepsilon_{\mu}$, which are (are not)
included in the effective $M \times M$ GUE Hamiltonian.
In the former case, the weight of ${\mb{o}}_{\nu}$ is 
$\sim M^{-1}$, whereas in the latter case it is $\sim \alpha^2 h^2/
N |\varepsilon_{\mu} - \varepsilon_{\nu}|^2$, where
$h^2 = N \Delta^2/\pi^2$ is the mean square of an element of 
%the perturbing matrix 
$H_{\rm U}$. Comparing the two estimates at
the energy difference $|\varepsilon_{\mu} - \varepsilon_{\nu}|
\sim M \Delta/2$ separating the two
regimes, we conclude $M \sim \alpha^2$, in agreement with
the exact result (\ref{eq:MN}).
\bibitem{renorm}
The last term in Eq.\ (\ref{eq:Uavg}) also vanishes in the GOE
once the renormalization of the local interaction from virtual
excitations is taken into account, see, e.g., Ref.\ \onlinecite{ABG}.
\end{references}
\end{document}